\documentclass[aps,prl,twocolumn,showpacs,groupedaddress]{revtex4}
\usepackage{graphicx}
\usepackage{color} 
\usepackage{graphicx}
\usepackage{epsfig}
\preprint{BSLCOgap/2008}
\begin{document}
\DeclareGraphicsExtensions{.eps, .jpg}
\bibliographystyle{prsty}
\input epsf

\title{Far-infrared absorption and the metal-to-insulator transition in  hole-doped cuprates} 
\author{S. Lupi$^{1}$, D. Nicoletti$^{1}$,  O. Limaj$^{1}$, L. Baldassarre$^{1}$, M. Ortolani$^{1}$, S. Ono$^{2}$, Yoichi Ando$^{3}$, and P. Calvani$^{1}$}
\affiliation {$^{1}$CNR-INFM Coherentia and Dipartimento di Fisica, Universit\`a di Roma "La Sapienza",
Piazzale A. Moro 2, I-00185 Roma, Italy}
\affiliation {$^{2}$Central Research Institute of Electric Power Industry, Komae, Tokyo 201-8511, Japan}
\affiliation {$^{3}$Institute of Scientific and Industrial Research, Osaka University, Ibaraki, Osaka 567-0047, Japan}

\date{\today}

\begin{abstract} 
By studying the optical conductivity of  Bi$_2$Sr$_{2-x}$La$_x$CuO$_6$ and Y$_{0.97}$Ca$_{0.03}$Ba$_{2}$Cu$_{3}$O$_{6}$, we show that the metal-to-insulator transition (MIT) in these hole-doped cuprates 
is driven by the opening of a small gap at low $T$ in the far infrared. Its width is consistent with the observations of Angle-Resolved Photoemission Spectroscopy in other cuprates, along the  nodal line of the $k$-space.  The gap forms as the Drude term turns into a far-infrared absorption, whose peak  frequency   can be approximately predicted on the basis of  a Mott-like transition. Another band in the mid infrared softens with doping but is less sensitive to the MIT. 
\end{abstract}


\maketitle

The parent compounds of high-$T_c$ cuprates are half-filled antiferromagnetic (AF) insulators where the Coulomb repulsion opens a wide charge-transfer (CT) gap ($\agt$ 1.5 eV) in the excitation spectrum. By adding a few percent  holes ($p$) or electrons ($n$) per Cu ion, the CT gap is filled by a broad infrared absorption, while Angle-Resolved Photoemission Spectroscopy (ARPES) reveals fully gapped single particle excitations at low temperature \cite{ShenARPES}.  
At $p \agt$ 0.05, like in La$_{2-x}$Sr$_x$CuO$_4$ (LSCO), or $n \agt$ 0.12 like in Nd$_{2-x}$Ce$_x$CuO$_4$ (NCCO), the insulator eventually turns at low $T$ into a superconductor, above $T_c$ into a "strange metal" characterized by a pseudogap in the density of states \cite{Timusk99}. Much effort has been devoted in the last two decades to understand these three phases and their excitation spectrum. However, less attention has been paid to the mechanism of the transition from the AF insulator to the strange metal and vice versa, which is far from being clear.  For example, the onset of a metallic state is observed at $p$ or $n$  values much higher than those which destroy the AF long range order \cite{Iye92}. Therefore, magnetism should not play a dominant role in the metal-to-insulator transition (MIT). 
The mechanisms responsible for the localization of the carriers then may be disorder due to dopant 
ions \cite{Atkinson}, local electron-spin interactions \cite{Defilippis}, or the electron-phonon coupling \cite{Cimento}.  

In the present paper we investigate the MIT by studying the behavior of the optical conductivity $\sigma_{1}(\omega)$ across the transition, in the single Cu-O layer cuprate Bi$_2$Sr$_{2-x}$La$_x$CuO$_6$ (BSLCO). 
Therein,   hole doping can be accurately controlled like in LSCO by replacing Sr by La. In BSLCO, $p$ decreases \cite{Ono00} for increasing $x$ according to a relation which is not linear, due to compensation effects, but well known \cite{Ando00, Ono03}. Finally, $T_c$ is low enough ($\simeq$ 30 K at optimum doping) to allow for a study of the normal state of the Cu-O planes at low $T$.  Here we have measured four single crystals \cite{Ono03} with $x$ = 1.0, 0.9, 0.8, and 0.7, corresponding to $p \simeq$ 0.03, 0.07, 0.10, and 0.12, respectively. As at $x=1.0$ one reaches the lowest $p\simeq 0.03$ allowed by the miscibility of La in the Bi$_2$Sr$_{2}$CuO$_6$ matrix. Therefore, we have also measured a single crystal of Y$_{1-x}$Ca$_{x}$Ba$_{2}$Cu$_{3}$O$_{6}$ (YCBCO) with $x$ = 0.03 and $p \simeq$ 0.015. As in YCBCO the CuO chains are empty and only the Cu-O plane contributes to the $ab$ optical response  \cite{Erb}, we could thus probe five Cu-O planes with decreasing hole doping, from an underdoped metallic state ($p \simeq$ 0.12) to an AF insulating phase ($p \simeq$ 0.015). 


\begin{figure}[!b]   \begin{center}  
\leavevmode    
\epsfxsize=8.6cm \epsfbox {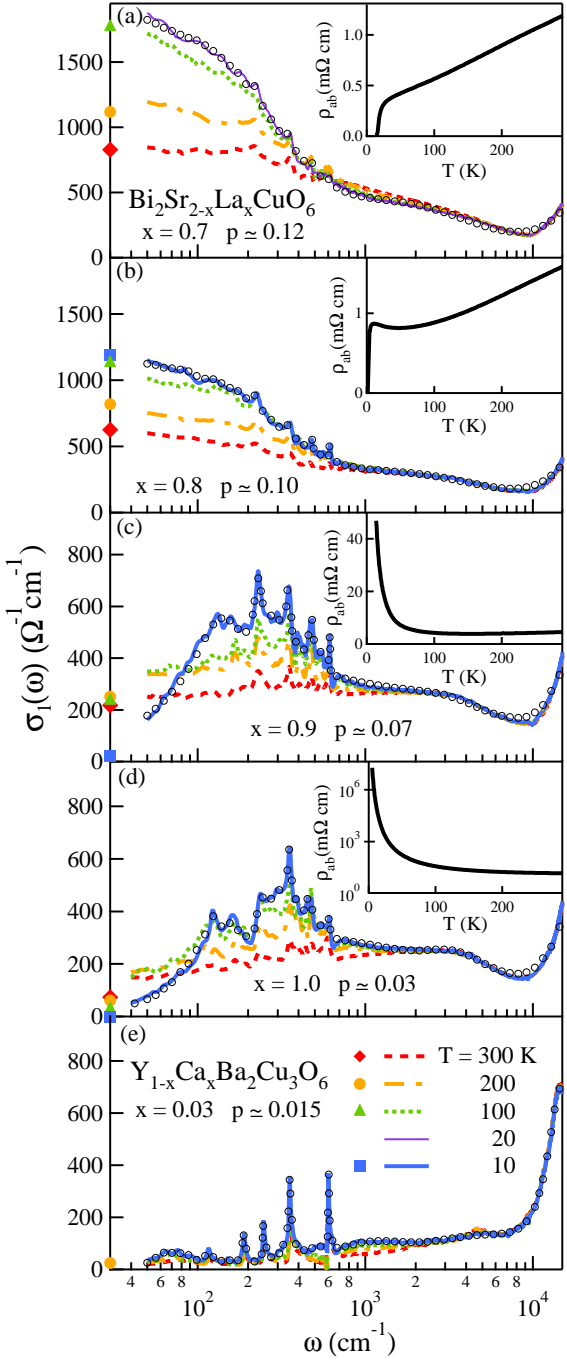}    
\caption{(Color online) Real part of the optical conductivity of Bi$_{2}$Sr$_{2-x}$La$_{x}$CuO$_{6}$ with different hole doping $p$, and of  Y$_{0.97}$Ca$_{0.03}$Ba$_{2}$Cu$_{3}$O$_{6}$, at selected temperatures. The circles show examples of the fits used to subtract the phonon contributions in Fig. 2. The symbols on the vertical axis indicate the dc conductivity measured in the same samples at the same $T$ ($\sigma_{dc}$ at 20 K is out of the scale of panel $a$). Full $ab$-plane resistivity curves are reported for BSLCO in the insets. The $\sigma_{dc}$ of YCBCO was measured at 200 K only.}
\label{sigma}
\end{center}
\end{figure}

The $ab$-plane resistivity $\rho_{ab}(T)$ of the four BSLCO crystals is reported in the insets of Fig. \ref{sigma} and dc conductivity values are shown, for selected temperatures, on the vertical axis of the same Figure. In $a$, for $p\simeq$ 0.12, $\rho_{ab}(T)$ shows a metallic behavior  above the superconducting transition at $T_c =$ 13 K.
The crystal with $p \simeq$ 0.10 in Fig. \ref{sigma}-$b$ displays a metallic behavior above 50 K, and a broad minimum above the transition to a superconducting state \cite{Ono03} at $T_c \simeq$ 1.4 K. At $p\simeq 0.07$, $\rho_{ab}(T)$ in $c$ is nearly constant down to $\sim$ 50 K. Below, it diverges for $T \to$ 0 according to a variable range  hopping regime  \cite{Ono03}.  A clearly semiconducting behavior at any $T$ ($d\rho_{ab}/dT<0$) is instead shown by the compound with $p\simeq$ 0.03 in Fig. \ref{sigma}-$d$. Therefore the MIT can be placed between $p\simeq 0.10$ and $p\simeq0.07$. This finding is consistent with the Mott-Ioffe-Regel limit, which fixes the metal-to-insulator crossover and, for the  Cu-O planes, can be written as \cite{Ando08} 
$k_F \cdot l = (h c_0 / e^2 \rho_{ab}) \sim 1$. Here, $k_F$ is the Fermi wavevector, $l$ the carrier mean free path, and $c_0$ is the $c$-axis lattice spacing. Indeed, following Ref. \onlinecite{Ono00}, from the $\rho_{ab}$ (10 K) in Fig. \ref{sigma} one obtains  $k_F \cdot l $ = 3.4 at $p$ = 0.10, $k_F \cdot l $ = 0.05 at $p$ = 0.07. Magnetic fields  on the order of 60 T displace the MIT  \cite{Ono00} to $p \simeq$ 1/8.

The $ab$-plane reflectivity $R(\omega)$  of the five samples was measured at near-normal incidence from 40 or 50 to 22000 cm$^{-1}$ at different $T > T_c$, shortly after cleaving the sample. The real part of the optical conductivity $\sigma_{1}(\omega)$, as obtained from $R(\omega)$ via Kramers-Kronig (KK) transformations,  is shown in Fig. \ref{sigma}. The extrapolations to high frequency were based on the data of Ref. \onlinecite{Terasaki}, those to zero frequency on Drude-Lorentz fits, which provided deviations from the measured $\sigma_{dc}(T)$  of a few percent. Afterwards, the extrapolations were adjusted to $\sigma_{dc}(T)$ (including that at 20 K in $a$, not shown).  In insulating YCBCO, a check value of $\sigma_{dc}$ was  measured at 200 K \cite{private} and reported in  Fig. \ref{sigma}-$e$. Figures \ref{sigma}-$a$ ($p \simeq$ 0.12) and  -$b$ ($p \simeq$ 0.10) exhibit a Drude term which partially shields the phonon peaks. These are instead well evident in panels $c$ ($p \simeq$ 0.07), $d$ ($p \simeq$ 0.03), and $e$ ($p \simeq$ 0.015). Their frequencies are in good agreement with those previously measured  \cite{TajimaPhon} on Bi$_2$SrLaCuO$_y$ and YBa$_{2}$Cu$_{3}$O$_{y}$. Around 10000 cm$^{-1}$, $\sigma_{1}(\omega)$ increases steeply due to the CT transition between Cu $3d$ and O $2p$ orbitals, as in the other cuprates.

\begin{figure}[!t]   \begin{center}  
\leavevmode    
\epsfxsize=8.6cm \epsfbox {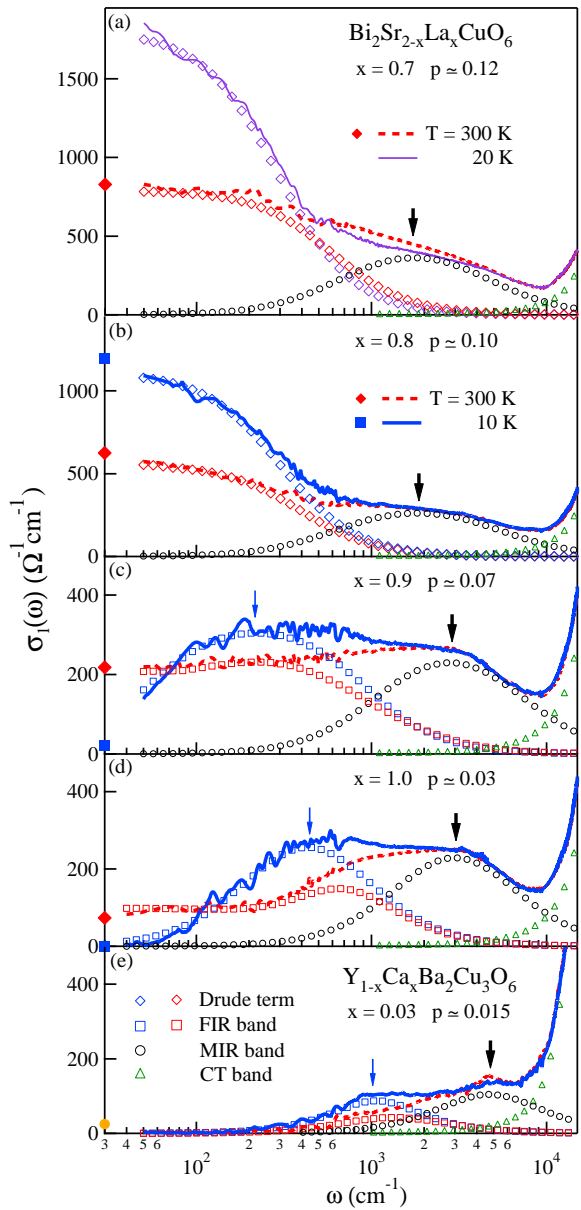}    
\caption{(Color online) Real part of the optical conductivity 
of the  Bi$_{2}$Sr$_{2-x}$La$_{x}$CuO$_{6}$ and  Y$_{0.97}$Ca$_{0.03}$Ba$_{2}$Cu$_{3}$O$_{6}$ crystals at two temperatures, after subtraction of the phonon contributions via Lorentzian fits (solid and dashed lines). 
Open symbols indicate the Lorentzian contributions listed in panel $e$, which were identified by fits to the subtracted data (not shown for clarity). In $c$ and $d$, the FIR band at 300 K includes a weak term peaked at $\omega$ = 0.  MIR and CT bands do not change with temperature. Thin and thick arrows indicate  $\omega_{FIR}$ and $\omega_{MIR}$, respectively. The dc values are the same as in Fig. 1.}
\label{subtract}
\end{center}
\end{figure}

Let us now focus on the broad far- and mid-infrared contributions in Fig. \ref{sigma}. To better understand their behavior with doping and temperature, in Fig. \ref{subtract} $\sigma_1(\omega)$ is reported at 300 K (dashed lines) and at the lowest $T$ (solid lines), after the phonon lines have been subtracted by accurate Lorentzian fits (those at the lowest T are shown in Fig. \ref{sigma}). 
Further fits - not reported in Fig. 2  as they could be hardly  distinguished from  the data - were then made on the subtracted spectra, by using a Drude term and two broad bands at $\omega_{FIR}$ and $\omega_{MIR}$ (open symbols).
In the metallic phase (Fig. \ref{subtract}-$a$ and -$b$)  at all temperatures, data are fit by a simple Drude term plus a $T$-independent mid-infrared (MIR) band peaked at $\omega_{MIR} \sim$ 2000 cm$^{-1}$. As $p$ decreases below the critical hole content $p_{MIT}$ (Fig.  \ref{subtract}-$c$), the Drude term  turns at low $T$ into a far-infrared (FIR) band peaked at $\omega_{FIR} \sim$ 200 cm$^{-1}$. Correspondingly, $\sigma_{dc}$ becomes vanishingly small. At $p\simeq$ 0.03 (Fig.  \ref{subtract}-$d$), a gap $\sim$ 100 cm$^{-1}$  opens in $\sigma_1(\omega)$ at $T$ = 10 K. Meanwhile, $\omega_{FIR}$ shifts to $\sim$ 400  cm$^{-1}$. Finally, at $p\simeq$ 0.015 in Fig. \ref{subtract}-$e$, the insulating gap extends along the whole far-infrared region, and the FIR peak is displaced to $\sim$ 1000 cm$^{-1}$. At 300 K, both in $c$ and $d$ the FIR band includes a background - suggestive of incoherent charge transport - which extends to $\omega$ = 0 and  accounts for the residual dc conductivity of these samples at room temperature. 


\begin{figure}[!t]   \begin{center}  
\leavevmode    
\epsfxsize=8.6cm \epsfbox {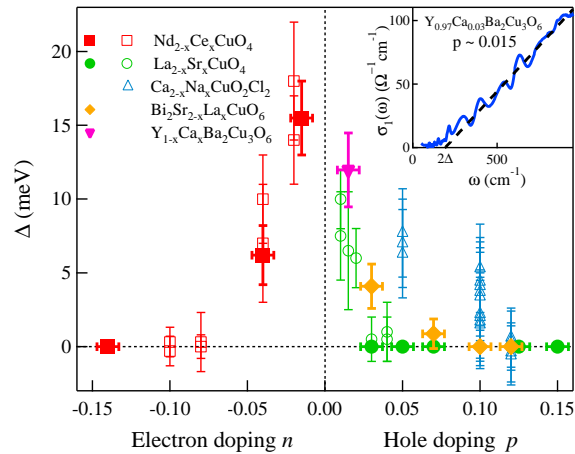}    
\caption{(Color online) Comparison between the gap $\Delta$ detected in 
several insulating cuprates  by ARPES \cite{ShenARPES} (open symbols) and by IR spectroscopy  (full symbols). Data on BSLCO and YCBCO are extracted from Fig. 2, those on NCCO from Ref. \onlinecite{Lupi99}, and on LSCO from Ref. \onlinecite{Lucarelli03}, by linearly extrapolating the gap edge to $\sigma_1$ = 0 (see an example in the inset).}
\label{ARPES-IR}
\end{center}
\end{figure}

The results of Fig. \ref{subtract} can be compared with those of ARPES experiments. 
Figure \ref{ARPES-IR} reproduces that in Ref.  \onlinecite{ShenARPES}, which shows the gap $\Delta$ at low doping and low $T$ in the density of states of different cuprates, measured at the leading edge midpoint along the   nodal line of the Brillouin zone. In BSLCO  with $p \simeq$ 0.05 and $p \simeq$ 0.07, a low-$T$ suppression of the spectral-weight was also observed along that direction in the $k$-space, but the gap width could not be measured for lack of resolving power  \cite{HashimotoARPES}. We have also plotted in Fig. \ref{ARPES-IR} both the present determinations of the gap in BSLCO, YCBCO, and those extracted from previous 
measurements in LSCO  \cite{Lucarelli03} and in NCCO \cite{Lupi99}. In all cases, the infrared gap 2$\Delta$ is  obtained after phonon subtraction, by a linear extrapolation to zero  \cite{Katsufuji} of the optical conductivity along the gap edge (see an example in the inset),  
 and then divided by two (by assuming full hole-electron symmetry). If also the IR gap were taken at mid-height, the full symbols would displace toward high frequencies by about a factor of 2.
In Fig. \ref{ARPES-IR},  independently of the procedure employed, both the infrared and  ARPES observations  consistently  indicate that a small gap -  just a few meV wide - opens in the density of states at the MIT. In LSCO, however, for reasons which deserve further investigations, the ARPES gap is not observed  the infrared  \cite{Lucarelli03,Padilla}.  In turn, the  band at $\omega_{MIR}$ can be associated with the incoherent background observed at $\sim 0.5$ eV below the quasi-particle peak in the antinodal ARPES spectra of cuprates 
\cite{ShenARPES,HashimotoARPES,MengCondMat}.   

The mechanism of the MIT in the Cu-O plane appears clearly in Fig.  \ref{subtract}: as $p$ decreases below  $p_{MIT}$, the carriers increasingly localize at low $T$ into states having an optical ionization energy $E_0$ measured by  $\omega_{FIR}$. $E_0$ increases  for decreasing $p$ and reaches about 0.12 eV in the limit of high dilution (Fig. \ref{sigma}-$e$). This value is in excellent agreement with previous observations in electron-doped NCCO at the lowest doping \cite{Calvani94}.

A simple calculation can predict $E_0$. Indeed, as usually done in semiconductors  \cite{Mott}, one may model the charge injected in the Cu-O plane as a hole orbiting with radius $R$ either around a defect or an impurity (in case of disorder) or within an attractive potential well due to lattice distortion (in a polaronic model).  $E_0$  is related to $R$ through the hydrogen-like equation

\begin{equation}
E_{0}=\left ( \frac{3}{2}\frac{a_{0}}{R} \right )^{2}Ry
\label{E0}
\end{equation}

\noindent
where $a_0$ = 0.0529 nm is the Bohr radius and $Ry$ = 109737 cm$^{-1}$ is  the Rydberg constant. 
The Mott transition occurs  at a 2-D critical density of holes per Cu ions $p_{MIT}$ such that \cite{nota}  the orbits of adjacent charges overlap in the Cu-O planes. This condition implies that the hole density $\rho_{MIT} \sim \frac{1}{\pi R^2}$ and the Cu density  $\rho_{Cu} =1/a^2$  satisfy the relation 

\begin{equation}
p_{MIT} = \rho_{MIT}/ \rho_{Cu} \sim \frac{a^2}{\pi R^2} 
\label{pmit}
\end{equation}

\noindent
Here, we use $a$ = 0.386 nm, averaged between  the Cu-O lattice parameters of \cite{TajimaPhon}  Bi$_2$SrLaCuO$_y$  and YBa$_2$Cu$_3$O$_y$ (0.383 nm  and 0.389 nm, respectively). Assuming from Fig. 2 $p_{MIT} \simeq 0.08$, and an average $\rho_{Cu}$ = 6.7 nm$^{-2}$, one obtains, from Eq. 2, $R \simeq 0.77$ nm and, from Eq. 1, $E_0 \simeq 1150$ cm$^{-1}$. This value is in very good agreement with the $\omega_{FIR} \simeq 1000$ cm$^{-1}$ observed here at high hole dilution (in the YCBCO with $p\simeq 0.015$ of Fig. \ref{subtract}-$e$).

Therefore, one should  find the origin of the FIR band in order to understand the MIT mechanism. A first candidate is disorder. In cuprates added with Zn impurities, or irradiated by high-energy particles, the metallic phase can be destroyed and the Drude spectral weight strongly reduced, due to a poor screening of the impurities and to the resulting  fluctuating potentials in the Cu-O planes \cite{Basov98, Basov94}. Recent calculations \cite{Atkinson}  show that such disorder effects are amplified in a $d$-wave electronic symmetry and that,  for increasing impurity content, the Drude term turns into a FIR peak at a finite frequency. This framework is consistent with the behavior of BSLCO, where the insulating phase  is reached by adding La impurities, not with that of LSCO which is opposite. It seems then that the  parameter which governs the metal-to-insulator transition is $p$, rather than $x$.

A different scenario can then be invoked to explain the MIT. Indeed, in BSLCO the FIR peak behaves with $p$ like a FIR band of NCCO vs. $n$ \cite{Lupi99}. This absorption was attributed to large polarons \cite{Cimento} and its softening was explained in terms of polaron-polaron interactions which  increase with $n$ \cite{Devreese, Lorenzana}.  
At room temperature, where $k_BT \approx \Delta$, incoherent polaron hopping takes place: this may explain the flat background observed in Fig. \ref{subtract}-$c$ and -$d$ and the resulting, non-vanishing $\sigma_{dc}$ at $p < p_{MIT}$. 
The above scenario is also consistent with recent calculations of the optical conductivity in hole-doped cuprates. They are based on a $t$-$J$-Holstein approach, where the FIR band has a dominant electron-phonon character, while the MIR band is attributed mainly to electron-spin interactions \cite{Mishchenko08}. The observation of the MIR band at $p > p_{MIT}$   therefore indicates that  local antiferromagnetic fluctuations survive in the underdoped metallic state, as also reported previously for YBCO \cite{Basov}. As $p$ decreases, in Fig. \ref{subtract} $\omega_{MIR}$ shifts steadily to higher energies, to reach $\sim$ 4500 cm$^{-1}$ at $p \simeq$ 0.015. This   value is consistent with the determinations of the MIR peak in the other cuprates ($\sim$ 0.5 eV) \cite{TimuskBasov}. Despite that shift, in Fig.  \ref{subtract} the MIR absorption does not seem to play a major role in the MIT. 


In conclusion, we have measured  both the dc and infrared conductivity of  BSLCO with different doping, and of a lightly doped sample of  YCBCO. We have thus monitored the mechanism of the metal-to-insulator transition in the hole-doped Cu-O plane.  For any  $p$ from 0.12 down to 0.015 we have detected a MIR band which hardens for decreasing $p$ but is poorly sensitive to the MIT. In the far infrared, on the contrary, the Drude term of the metallic phase collapses at the MIT into a FIR band at finite frequency. As $p$ is further decreased, this band shifts to higher energies, leaving behind a gap  at low $T$, a flat absorption tail at high $T$. The FIR gap here observed is in quantitative agreement with that reported by ARPES in other cuprates  with similar doping at low $T$, and the FIR peak frequency is correctly predicted by a simple Mott-transition model. 
Both bands here identified are consistent with a recent  model, where the carrier  localization is interpreted in terms of charge and spin polarons. According to this interpretation, the FIR band is mainly due to electron-phonon interaction,  the MIR band to electron-spin interaction.

We are gratefully indebted to Andreas Erb for providing the YCBCO crystal. This work has been partially funded by  PRIN 2005022492. S.O. was supported by KAKENHI No. 20740213.



\end{document}